\documentclass{ws-procs975x65}

\begin{document}

\title{Generalisations of the Einstein--Straus 
model to cylindrically symmetric settings}

\author{Filipe C. Mena$^{1,2}$, Reza Tavakol$^2$ and Ra\"ul Vera$^3$}

\address{$^1$Departamento de Matem\'atica, Universidade do Minho, 4710 Braga, Portugal\\ 
$^2$Astronomy Unit, Queen Mary, University of London, London E1 4NS, U.K.\\
$^3$School of Mathematical Sciences, Dublin City University, Glasnevin, Dublin 9, Ireland} 

\maketitle
\abstracts{We study generalisations of the Einstein--Straus
model in cylindrically symmetric settings by considering the 
matching of a static space-time to a non-static 
spatially homogeneous space-time, preserving the symmetry. We find that such models 
possess severe restrictions, such as constancy of one of the metric coefficients
in the non-static part. 
A consequence of this is that it is impossible to embed static locally cylindrically
symmetric objects in reasonable spatially homogeneous cosmologies.}
\section{Introduction}
Possible influences of the large scale evolution of the universe on local dynamics 
has been the subject of intense study in cosmology for a long time.
As an attempt in answering this question,
Einstein and Straus\cite{Einstein-Straus} considered a model consisting of the matching
of an inner Schwarzshild metric, approximating
the local neighbourhood in the solar system, and an outer 
dust Friedmann-Lemaitre-Robertson-Walker (FLRW),
representing the expanding universe.
They showed that such matching was possible across any comoving 
3-sphere, provided the total mass inside the 3-sphere was equal to the
Schwarzshild mass contained in it. Thus according to this model
the global expansion of the universe exerts no influence on 
the vacuum region surrounding the Schwarzshild mass.

Given the potential importance of this result, it is important
to ask whether it still holds if some of the inevitable idealisations involved
in the Einstein--Straus model
are relaxed.
A number of attempts have been  made to generalise
the original setting, by including other source
fields and FLRW geometries\cite{Seno-Vera}.
For example, it has been shown that in 
order to match a static spacetime to an
expanding FLRW the configuration has to be "almost spherical"\cite{Mars01}.
This indicates that the 
original Einstein-Straus model is unstable.
Realistic cosmological models, however, cannot be expected to be exactly
homogeneous and isotropic, so it would be interesting to study modified 
Einstein-Straus models which include anisotropic cosmologies.

Here we summarize and extend recent results which allow anisotropies
in the Einstein-Straus model\cite{mtv}. This is done by 
studying embeddings of static cylindrically symmetric 
cavities in expanding homogeneous and anisotropic universes. 
\section{Matching Conditions}
We consider the matching between two locally cylindrically symmetric spacetimes,
preserving the symmetry\cite{Vera01}. 
The matching conditions between
two spacetimes $(g^+,M^+)$ and $(g^-,M^-)$
across a $3-$hypersurface $\sigma$
are twofold\cite{maseaxconf}: 
firstly the equality of the 
first fundamental forms at $\sigma$
\begin{equation}
\label{first}
\bar g^{\pm}_{ab}=g^{\pm}_{\alpha\beta} e_{a}^{\pm\alpha} e_{b}^{\pm\beta}|_\sigma
\end{equation}
and secondly the equality of the generalised second fundamental forms at $\sigma$
\begin{equation}
\label{second}
H^{\pm}_{ab} =-\ell^{\pm}_{\alpha}e^{\pm\beta}_{a}\nabla^{\pm}_{\beta}
e^{\pm\alpha}_{b}|_\sigma,
\end{equation}
where $\ell^\pm_\alpha$ are the rigging forms\cite{mtv}. 
These conditions imply, in turn,
the Israel conditions which are the equality at $\sigma$ of
\begin{equation}
\label{isra}
S^{\pm}_{\beta}=n^{\pm\alpha} T^{\pm}_{\alpha\beta},
\end{equation}
where $n^{\pm \alpha}$ are the normal vectors to $\sigma$.

Here we take $(g^-,M^-)$ to be the most general cylindrically symmetric static metric
\begin{equation}
\label{static}
d s^{2-}=-A^{2} dT^{2} + B^2 d\rho^{2} 
+C^{2} d\tilde\phi^{2} + D^{2} d\tilde z^{2}+ 2Ed\tilde \phi d\tilde z,
\end{equation}
where $A,B,C,D,E$ are functions of $\rho$.

\noindent For $(g^+,M^+)$ we take the most general spatially homogeneous spacetimes which, in order
to have a globally defined axis, are required to have a $G_4$ on $S_3$. 
These can be cast in a metric form adapted to the killing vectors 
$\partial_\phi$ and $\partial_z$: 
\begin{equation}
\label{compact}
ds^{2+}=-\hat A^2dt^2+\hat B^2dr^2-2\epsilon r\hat B^2drdz+\hat C^2d\phi^2 +2\hat Edz d\phi
+\hat D^2dz^2,
\end{equation}
with 
\begin{eqnarray}
\hat A^2&=&1;~~~~\hat B^2=b^2(t);~~~~\hat C^2=b^2(t)\Sigma^2(r)+n a^2(t)(F(r)+k)^2,\nonumber\\
\hat D^2&=&a^2(t)+\epsilon r^2b^2(t);~~~~\hat E=na^2(t)(F(r)+k),\nonumber
\end{eqnarray}  
where $\epsilon=0,1~~n=0,1~~\epsilon n=\epsilon k=0$ and $\Sigma(r)$ and $F(r)$ depend 
on $k=0,-1,1$. Note that these metrics include all Bianchi types. 

The matching conditions (\ref{first}) then imply\cite{mtv} 
\begin{equation}
\label{pre1}
\hat D \stackrel{\sigma}{=} D;~~~~~
\hat C \stackrel{\sigma}{=} C;~~~~~
\hat E \stackrel{\sigma}{=} E.
\end{equation}
From (\ref{second}) we obtain, in particular, 
\begin{equation}
\label{exterior}
\hat{D}_{,t}\hat{C}_{,r}-\hat{D}_{,r}\hat{C}_{,t}\stackrel{\sigma}{=}0,~~~
\hat{E}_{,t}\hat{D}_{,r}-\hat{E}_{,r}\hat{D}_{,t}
\stackrel{\sigma}{=}0,~~~
\hat{E}_{,t}\hat{C}_{,r}-\hat{E}_{,r}\hat{C}_{,t}\stackrel{\sigma}{=}0.
\end{equation} 
from which we get $n=0$. 
\section{Results} 
Assuming that the locally cylindrically 
symmetric (LSC) static regions are spatially compact and simply connected,
and using $n=0$ together with the matching conditions, we obtain the following results:
\vspace{0.1cm}

\noindent {\bf Proposition:}
{\em A non-static Bianchi II, VIII or IX spacetime cannot
be matched to a LCS static region across a non-spacelike hypersurface preserving the
symmetry.}
\vspace{0.2cm}

\noindent
{\bf Theorem:}
{\em The only possible non-static spatially homogeneous spacetimes
that can be matched to a
LCS static region
across a non-space-like hypersurface preserving
the symmetry are given by
\begin{equation}
\label{exteriords2}
ds^2=-dt^2+\beta^2 dz^2
+b^2(t)\left[(dr-\epsilon r dz)^2+
\Sigma^2(r,k) d\varphi^2\right],
\end{equation}
where $\beta$ is a constant
and $\epsilon=0,1$ is such that $\epsilon k=0$.}
\vspace{0.2cm}

\noindent
As a consequence of this theorem the only non-zero fluid expansion components are
\begin{equation}
\theta_{11}=\theta_{22}=\frac{b_{,t}}{b}.
\end{equation}
The vanishing of $\theta_{33}$ is a severe constraint as
far as cosmologically interesting models are concerned since there
is no expansion along one spatial direction (spanned by $\partial_z+\epsilon r \partial _r$).
Using
the Israel conditions (\ref{isra}) we also prove the following corollaries:
\vspace{0.2cm}

\noindent  
{\bf Corollary 1 (perfect fluid):}
{\em If the non-static spacetime has a perfect-fluid 
with $\rho+p>0$
then it must have metric 
(\ref{exteriords2}) with $\epsilon=0,~b(t)=\sqrt{\alpha t-kt^2},~\alpha>0$ and }
\begin{equation}
\label{stiff}
\rho=p=\frac{\alpha^2}{4 t^2(\alpha-k t)^2}.
\end{equation}
{\bf Corollary 2 (vacuum):}
{\em If the static spacetime is vacuum then the non-static spacetime is also vacuum.} 
\vspace{0.2cm}

\noindent
The above results show that 
there are no evolving perfect fluid
Bianchi spacetimes with $\rho\neq p$ satisfying
the dominant energy condition and containing
a locally cylindrically symmetric static cavity. They therefore demonstrate  
that the Einstein--Straus result cannot be generalised in this way
to cylindrical symmetry.
\vspace{0.2cm}

\noindent
{\bf Acknowledgments}
\vspace{0.2cm}

\noindent
FCM thanks British
Council/CRUP for grant N$13/03$, Funda\c{c}\~ao Calouste Gulbenkian for
grant 21-58348-B and Centro de Matem\'atica, Univ. Minho, for support.

\end{document}